\documentclass[secnumarabic,amssymb, nobibnotes, pra, superscriptaddress, preprint]{revtex4-1}

\usepackage{comment}

\usepackage{graphicx}
\usepackage{dcolumn}
\usepackage{bm}
\usepackage{amssymb}

\usepackage{geometry}
\usepackage{longtable}
\usepackage{url}
\usepackage{booktabs}
\usepackage{float}

\begin{document}

\title{Contribution of high-$nl$ shells to electron-impact ionization process}

\author{V.~Jonauskas}%
\email[]{Valdas.Jonauskas@tfai.vu.lt}
\affiliation{Institute of Theoretical Physics and Astronomy, Vilnius
University, \ A. Go\v{s}tauto 12, LT-01108  Vilnius, Lithuania}

\author{A. Kynien\.{e}}%
\affiliation{Institute of Theoretical Physics and Astronomy, Vilnius
University, \ A. Go\v{s}tauto 12, LT-01108  Vilnius, Lithuania}

\author{G. Merkelis}%
\affiliation{Institute of Theoretical Physics and Astronomy, Vilnius
University, \ A. Go\v{s}tauto 12, LT-01108  Vilnius, Lithuania}

\author{G. Gaigalas}%
\affiliation{Institute of Theoretical Physics and Astronomy, Vilnius
University, \ A. Go\v{s}tauto 12, LT-01108  Vilnius, Lithuania}

\author{R. Kisielius}%
\affiliation{Institute of Theoretical Physics and Astronomy, Vilnius
University, \ A. Go\v{s}tauto 12, LT-01108  Vilnius, Lithuania}

\author{S. Ku\v{c}as}%
\affiliation{Institute of Theoretical Physics and Astronomy, Vilnius
University, \ A. Go\v{s}tauto 12, LT-01108  Vilnius, Lithuania}

\author{ \v{S}. Masys}%
\affiliation{Institute of Theoretical Physics and Astronomy, Vilnius
University, \ A. Go\v{s}tauto 12, LT-01108  Vilnius, Lithuania}

\author{ L. Rad\v{z}i\={u}t\.{e}}%
\affiliation{Institute of Theoretical Physics and Astronomy, Vilnius
University, \ A. Go\v{s}tauto 12, LT-01108  Vilnius, Lithuania}

\author{P. Rynkun}%
\affiliation{Institute of Theoretical Physics and Astronomy, Vilnius
University, \ A. Go\v{s}tauto 12, LT-01108  Vilnius, Lithuania}

\begin{abstract}

Contribution to electron-impact ionization cross sections from excitations to 
high-$nl$ shells and a consequent autoionization is investigated. We perform 
relativistic subconfiguration-average and detailed level-to-level calculations 
for this process. Ionization cross sections for the W$^{27+}$ ion are presented 
to illustrate large influence of the high shells ($n \geqslant 9$) and orbitals 
($l \geqslant 4$) in the excitation-autoionization process. The obtained results 
show that the excitations to the high shells ($n \geqslant 9$) increase cross 
sections of indirect ionization process by a factor of two compared to the 
excitations to the lower shells ($n \leqslant 8$). The excitations to the shells 
with the orbital quantum number $l=4$ give the largest contribution compared with 
the other orbital quantum numbers $l$. Radiative damping reduces the cross 
sections of the indirect process approximately two times in the case of the 
level-to-level calculations. Determined data show that the 
excitation-autoionization process contributes approximately $40\%$ to the total 
ionization cross sections.

\end{abstract}


\maketitle

\section{Introduction}

Energy losses from heavy elements, such as tungsten, due to radiative emission 
are one of the crucial problems to be overcome for the successful performance 
of themonuclear reactors.  Nevertheless, tungsten is used as the plasma-facing 
component in the modern fusion facilities because of its essential properties, 
like high-energy threshold for sputtering, low sputtering yield, and excellent 
thermal features. Even small concentration of tungsten ions ($\sim 10^{-4}$) 
relative to the electron density prevents ignition of a deuterium-tritium plasma 
\cite{Kallenbach_2005ppcf_47_b207}. Theoretical modeling provides information 
about processes in such harsh conditions. However, fusion plasma modeling 
requires a significant amount of atomic data. The ionization and recombination 
processes determine the charge-state distribution in plasma. The electron-impact 
single ionization is the strongest one among the ionization processes. 
Contribution of the double ionization as a rule is much weaker compared with 
the single ionization. Furthermore, study of the double ionization process is 
quite complicated \cite{2007jpb_40_r39_pindzola, 2012epjd_66_287_colgan, 
2014pra_89_052714_jonauskas}.

Electron-impact ionization for singly charged tungsten ions was previously 
studied by using the crossed-beam technique \cite{1984jpb_17_2707_montague}. 
These data were supplemented by the electron-impact single, double, and triple 
ionization measurements of W$^{q+}$ ions in the charge states $q=1-10$, $q=1-6$, 
and $q=1-4$, respectively \cite{1995jpb_28_2711_stenke, 1995jpb_28_4853_stenke}. 
The configuration-average distorted-wave (CADW) calculations provided good 
agreement with experimental measurements for higher ionization stages 
($q \geqslant 4 $) \cite{1997pra_56_1654_pindzola}. The DW approach succeeded 
in getting a very good agreement with experiment for the ionization stages up to 
W$^{9+}$ ion and therefore it  was  applied to all isonuclear sequence 
\cite{2005pra_72_052716_loch}. Further level-to-level studies 
\cite{2009ljp_49_415_jonauskas} included W atom and W$^{2+}$ ion in the 
binary-encounter-dipole model as well as W$^{4+}$ and $W^{6+}$ ions in the DW 
approach. For the tungsten isonuclear sequence, cross sections were calculated 
using semirelativistic and relativistic DW methods for configurations and 
subconfigurations, respectively \cite{2005pra_72_052716_loch}. Influence of the 
excitation-autoionization (EA) process, radiative damping, and relativistic 
effects were analyzed.  The work of Loch et al. \cite{2005pra_72_052716_loch} 
considered the electron-impact excitations to all shells with $n\leqslant 8$ 
and $l \leqslant 3$.  The investigation has determined that the EA contribution 
is relatively small compared to direct ionization (DI) for the W$^{11+}$ to 
W$^{27+}$ ions. However, later experimental measurements for the W$^{17+}$ ion 
have demonstrated a significant role of indirect process 
\cite{2011jpb_44_165202_rausch}. Theoretical study of cross sections using the 
Dirac-Fock-Slater approach confirmed these findings 
\cite{2014jpb_47_075202_zhang}. Furthermore, the authors found that Maxwellian 
rate coeffcients are larger than the CADW rate coefficients by about 16\%. 
The discrepancy was attributed to the EA channels originating from the high-$n$ 
shells  up to $n=38$. 

Therefore, the main aim of our work is to determine the influence of excitations 
to the high-$n$ shells which lead  to the EA process. For the study, we have 
chosen the W$^{27+}$ ion which has only one $4f$ electron in the valence shell. 
At first, we will present calculated results of the EA contribution for 
subconfigurations. Later, level-to-level distorted-wave (LLDW) data in a single 
configuration approach are analyzed. In addition, the influence of the radiative 
damping is studied in both cases. The contribution of the direct and indirect 
processes is compared for the W$^{27+}$ ion.

\section{Theoretical approach}

Direct and indirect processes contribute to the total electron-impact 
single-ionization cross sections from  the level $i$ of $A^{q+}$ ion to the 
level $j$ of $A^{(q+1)+}$ ion:
\begin{equation}
\sigma_{ij} (\varepsilon) = \sigma_{ij}^{\mathrm{DI}}(\varepsilon) + \sum_{k} \sigma_{ik}^{\mathrm{exc}}(\varepsilon) B_{kj},
\label{sigma}
\end{equation}
where $\sigma_{ij}^{\mathrm{DI}} (\varepsilon)$ is the direct ionization  cross section at the electron energy $\varepsilon$, $\sigma_{ik}^{\mathrm{exc}}$ is the electron-impact excitation cross section to the level $k$ of the $A^{q+}$ ion. Autoionization branching ratio $B_{kj}$ is determined by the expression:
\begin{equation}
B_{kj} = \frac{A_{kj}^{\mathrm{a}} + \sum_{n} A_{kn}^{\mathrm{r}} B_{nj}}{\sum_{m} A_{km}^{\mathrm{a}} + \sum_{n} A_{kn}^{\mathrm{r}}}, 
\label{branching}
\end{equation}
where $A^{a}$ and $A^{r}$ are the Auger and radiative transition probabilities, 
respectively.  Therefore, inclusion of the branching ratios in the cross section 
calculations leads to the radiative damping of the indirect process. The second 
term in the numerator presents the transition from the level $k$ of the initial 
ion to the level $j$ of the final ion through the intermediate levels $n$ of the 
initial ion reached by the radiative transitions. This term is not studied in 
the current work because the amount of calculations drastically increases. 
Thus, the indirect ionization consists of the two-step process: the excitation 
with subsequent autoionization. It is evident that the branching ratios for the 
levels below the ionization threshold are equal to zero. Furthermore, the 
branching ratios are equal to zero for the levels above the ionization threshold 
if they cannot decay through the Auger transitions directly or through the 
intermediate states. The total ionization cross section for the initial level 
$i$ is obtained by performing summation over the all final levels $j$ in 
Eq. (\ref{sigma}). 

We do not consider higher order indirect ionization process such as a 
resonant-excitation double-autoionization. Previous analysis of this process 
for the W$^{17+}$ ion determined a negligible contribution for the ionization 
cross sections \cite{2014jpb_47_075202_zhang}. 

The electron-impact excitation and ionization cross sections are determined in 
the DW approximation using the Flexible Atomic Code (FAC) 
\cite{2008cjp_86_675_Gu} which implements the Dirac-Fock-Slater method. 
For moderate and highly charged ions, the direct ionization is accurately 
described by the DW approximation. The ionization cross sections are calculated 
in the potential of the ionizing ion. Calculations in the potential of the 
ionized ion provide approximately 10\% smaller cross sections for the direct 
ionization.

\section{Results}

The ground configuration of the W$^{27+}$ ion is [Kr]$4d^{10}4f$ which consists 
of two levels. Our analysis of the ionization cross sections is based on the 
study of the ground level ($4f_{5/2}$). Cross sections from the first excited 
level ($4f_{7/2}$) are similar to the ones from the ground level and are not 
presented here. 

\begin{figure}
 \includegraphics[scale=0.4]{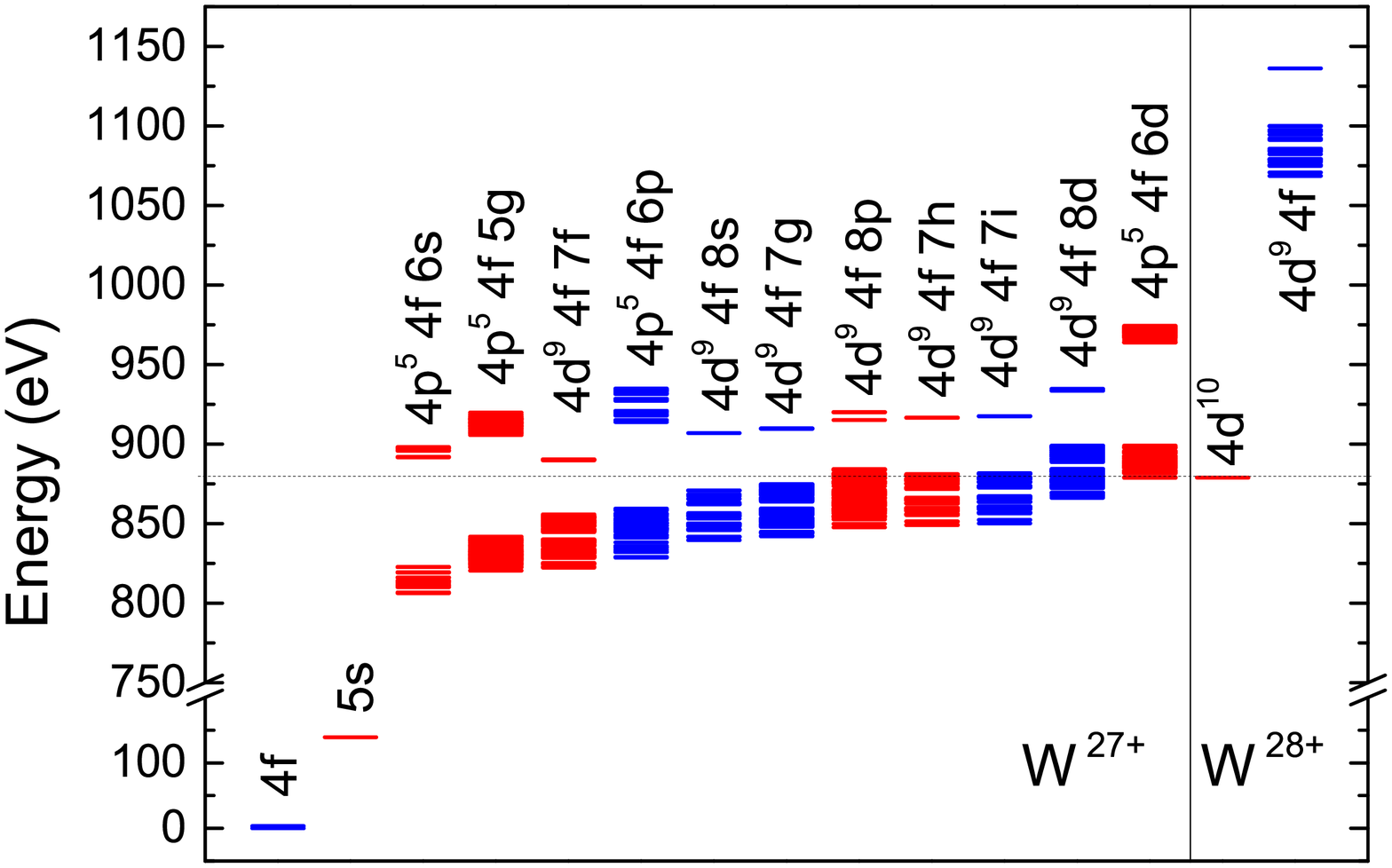}
 \caption{\label{energies} (Color online)  Energy levels of two lowest 
configurations and configurations which straddle the ionization threshold 
for the W$^{27+}$ ion. Red color -- even configurations, blue color -- odd 
configurations.} 
\end{figure}

The lowest two configurations of the W$^{27+}$ ion together with the 
configurations which energy levels straddle the ionization threshold are shown 
in Fig.~\ref{energies}. The presented configurations with the energy levels 
near the ionization threshold are produced by the one-electron promotions from 
the ground configuration of the W$^{27+}$ ion. They mainly correspond to the 
excitations from the $4d$ shell up to the shells with $n\leqslant8$. All of 
these configurations, except for the $4d^{9}4f8d$ one, have average energies 
below the ionization threshold. Thus, they do not provide contribution  to the 
ionization cross sections in the CADW approach. A few  configurations formed by 
promotion from the $4p$ shell also straddle the ionization threshold. However, 
the one-electron excitations reach only $n=6$ shell in this case. 

The ionization process affecting the $4s$, $4p$, $4d$, and $4f$ shells 
contribute to the DI process. On the other hand, study of the W$^{17+}$ ion has 
demonstrated that ionization from the $4s$ shell lies above the 
double-ionization threshold  and contributes to the indirect double-ionization 
process \cite{2014jpb_47_075202_zhang}. 

Table I  shows calculated energy levels for the lowest configurations of the 
W$^{27+}$, W$^{28+}$, and W$^{29+}$ ions. In addition, the lowest levels of the 
configurations for the W$^{28+}$ ion produced from the ground configuration of 
the W$^{27+}$ ion after an electron is ionized from the $3d$, $4s$, $4p$, or 
$4d$ shells are shown. One can see that the calculated ionization threshold 
energy equals to 878.89 eV. The ionization energy obtained from the scaled 
electron binding energies is given as $881.40 \pm 1.61$ eV 
\cite{2006adndt_92_457_kramida}. That is in quite close agreement with our 
determined value. The ionization of the long-living level of the [Kr]$4d^{10}5s$ 
configuration corresponds to threshold energy 739.56 eV. This configuration can 
decay to the ground configuration through the weak electric octupole transitions  
(lifetime $2.16 \cdot 10^{2}$ s). 

\begin{figure}
 \includegraphics[scale=0.4]{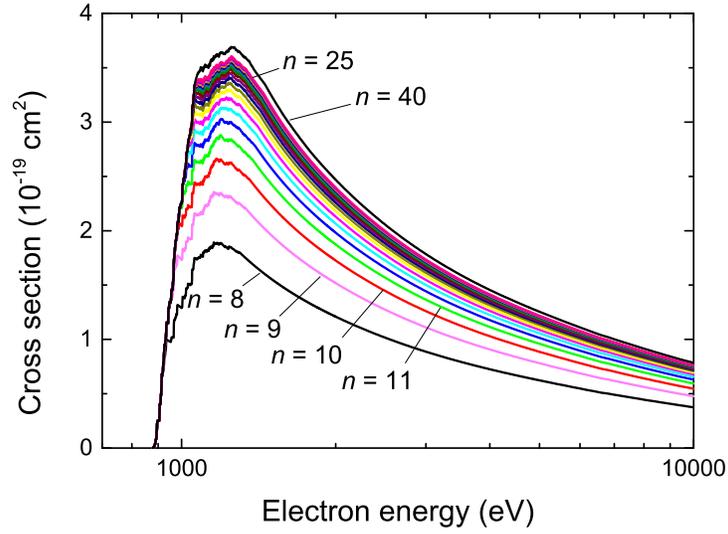}
 \caption{\label{ea} (Color online) EA channels to the high-$nl$ shells for the 
W$^{27+}$ ion. SCADW calculations. }
\end{figure}

\begin{figure}
 \includegraphics[scale=0.4]{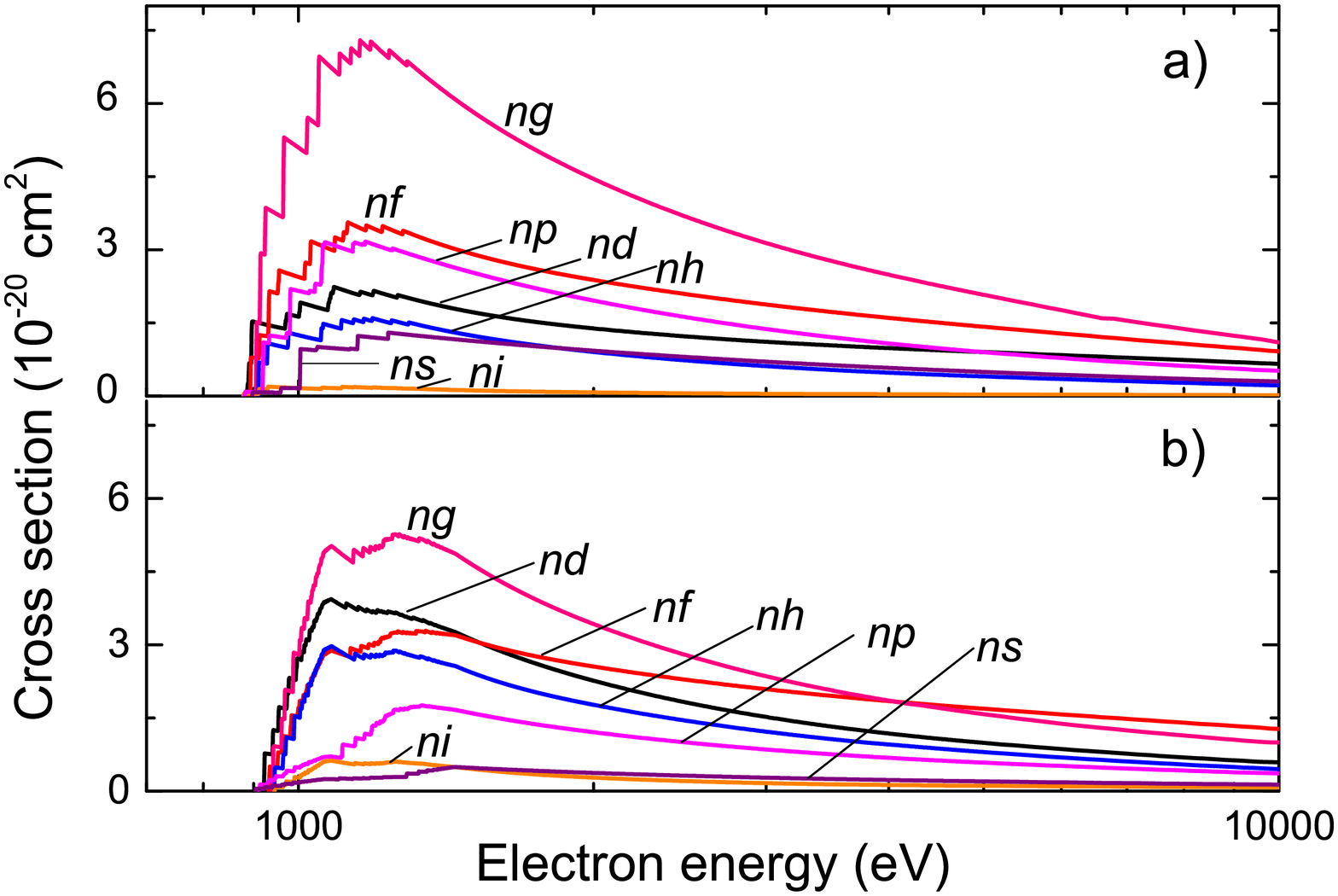}
 \caption{\label{ea2nl} (Color online) EA cross sections corresponding to 
excitations to various orbitals: a) $n \leqslant$ 8, b)  
$9 \leqslant n \leqslant 40$. SCADW calculations. }
\end{figure}

Figure~\ref{ea} shows contribution of the EA channels originating from the 
excitations to the high-$n$ shells up to $n=40$. These data have been produced 
using the subconfiguration-average distorted-wave (SCADW) approach.  As it was 
mentioned eralier, the previous study using the semirelativistic CADW 
calculations included only the excitations to the outer shells with 
$n \leqslant 8$ and $l \leqslant 3$ \cite{2005pra_72_052716_loch}. It can be 
seen from Fig.~\ref{ea} that the additional EA channels originating from the 
excitations to the higher shells increase cross sections for the indirect part 
approximately by factor of two. The convergence for the cross sections is 
reached when the excitations to the high-$n$ shells are included. Furthermore, 
our calculations take into account the excitations to the shells with 
$l \leqslant 6$.  Fig.~\ref{ea2nl} shows that the largest contribution to the 
EA process comes from the inner shell excitations to the $l=4$ orbital. The 
same trend is observed for the $n \leqslant 8$ and $9 \leqslant n \leqslant 25$ 
shells. However, the contribution to the EA process of the excitations to the 
$l=4$ orbital decreases if compared with other orbitals in the latter case. 
It is interesting that none of the excitations from the $4s$, $4p$, or $4d$ 
shells to the $l=4$ orbital give the largest cross sections. However, the total 
contribution of the excitations from the $4s$, $4p$, and $4d$ shells to the 
$l=4$ orbital is the largest compared to other orbitals. It is worth to note 
that excitations to $l=5$ orbital lead to the strong EA channel for which the 
relative contribution increases for $9 \leqslant n \leqslant 25$ shells 
(Fig.~\ref{ea2nl}b). For these shells, the largest increase of the relative 
contribution occurs for the excitations to a $l=2$ orbital compared to the 
$n \leqslant 8$ case. This can be explained by the fact that the strongest 
excitations to the $l=2$ orbital take place from the $4d$ shell. However, the 
configurations $4d^{9}4f\,nd$ with $n=5,6,7$ are below the ionization threshold, 
and they do not contribute to the EA process. Only some levels of the 
$4d^{9}4f\,8d$ configuration are above the ionization threshold 
(Fig.~\ref{energies}).  On the other hand, all levels of $4d^{9}4f\,8g$ 
configuration are above the ionization threshold. Furthermore, the energy levels 
of the $4d^{9}4f\,7g$ configuration straddle the ionization threshold. It 
explains much larger relative contribution to the EA process of the excitations 
to the $l=4$ orbital for $n \leqslant 8$ shells.

Other interesting result occurs when the EA channel originating from excitations 
to the $l=3$ orbital starts to dominate at high electron energies for the 
$9 \leqslant n \leqslant 25$ shells. Contribution of the EA channel to the $l=4$ 
orbital is slightly smaller at the higher energies when the excitations to the 
large principal quantum numbers $n$ are considered (Fig.~\ref{ea2nl}b).

\begin{figure}
 \includegraphics[scale=0.4]{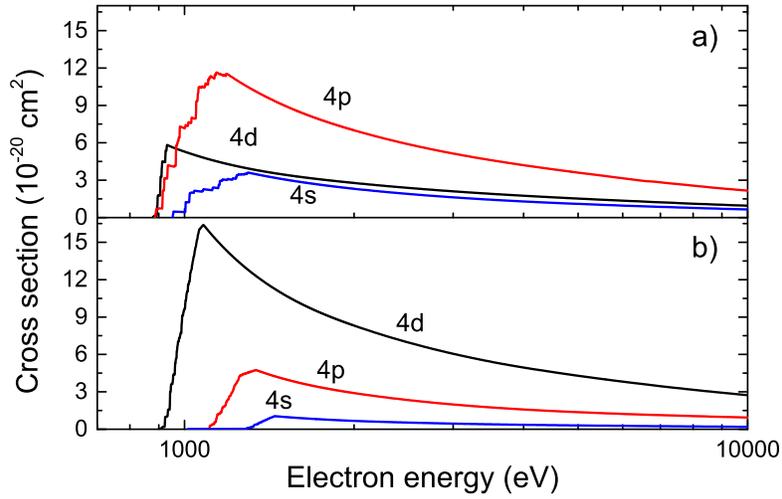}
 \caption{\label{ea4l} (Color online) EA cross sections corresponding to 
excitations from the $4s$, $4p$, and $4d$ shells: a) $n \leqslant$ 8, b) 
$9 \leqslant n \leqslant 40$. SCADW calculations. }
\end{figure}

The EA channel for the excitations from the $4d$ shell is predicted to produce 
the largest contribution compared to the EA channels from the $4s$ and $4p$ 
shells because the excitation cross sections are larger for the $4d$ shell. 
However, as it is mentioned above, many configurations produced by the promotion 
from the $4d$ shell to the $nl$ shell for $n \leqslant 8$ are below the 
ionization threshold. Therefore, these configurations do not contribute to the 
EA process. Figure \ref{ea4l}a demonstrates that the contribution of the 
excitations from the $4d$ shell to the EA process for $n \leqslant 8$ is more 
than two times smaller compared to the EA channel for the excitations from the 
$4p$ shell. The similar result was obtained for the W$^{17+}$ ion when the 
excitations only up to $n \leqslant 8$ were considered 
\cite{2014jpb_47_075202_zhang}. However, the situation drastically changes for 
the excitations from the $4d$ to the higher shells because all arising 
configurations are above the ionization threshold (Fig.~\ref{ea4l}b). 
Contribution of this EA channel is much larger compared to the EA channel from 
the $4p$ shell. Furthermore, the contribution of the EA channel from the $4d$ 
shell for the excitations to $9 \leqslant n \leqslant 25$ shells is comparable  
to the total cross sections of the EA process for the excitations to 
$n \leqslant 8$ shells (Fig.~\ref{ea4l}a). 

Calculations of the EA process for subconfigurations can differ from those for 
the detailed level-to-level studies. In our SCADW calculations, the energy 
levels are grouped into the subconfigurations. For the energy levels that 
straddle the ionization threshold, the corresponding energy of the 
subconfiguration can be above or below the ionization threshold. 
The subconfiguration with the energy above the ionization threshold contributes 
to the EA process. Therefore, all levels of the subconfiguration, even those 
located below the ionization threshold, are included when the cross sections 
are calculated. On the other hand, for the subconfiguration below the ionization 
threshold, the contributions from the corresponding energy levels above the 
ionization threshold are neglected. 

\begin{figure}
 \includegraphics[scale=0.4]{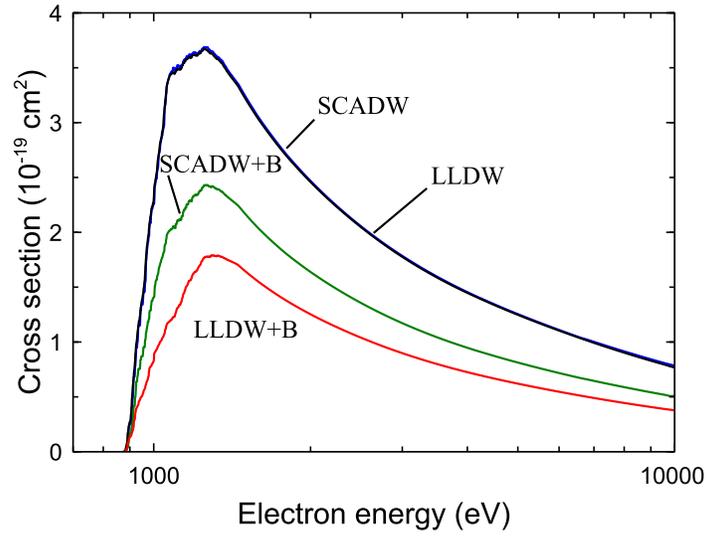}
 \caption{\label{damping} (Color online) LLDW and SCADW cross sections for the 
EA process without (LLDW: black, SCADW: blue) and with (LLDW+B: red, SCADW+B: 
green) radiative damping. }
\end{figure}

It is interesting to note that the EA cross sections obtained using SCADW and 
LLDW approaches are in good agreement when the autoionization branching ratios 
are not included in calculations (Fig.~\ref{damping}). Figure~\ref{damping} 
shows that the radiative damping has the large impact on the ionization cross 
sections. However, it was found in the previous study that the autoionization 
branching ratios are close to 1 for this ionization stage 
\cite{2005pra_72_052716_loch}. The difference can be explained by the fact that 
the previous study used the configuration-average quantities while the current 
SCADW calculations employ the subconfigurations. The higher values of the cross 
sections for the SCADW calculations compared to the LLDW results suggest that 
the autoionization branching ratios for the CADW data are closer to 1. 
Therefore, the radiative decay paths are restricted for some excited 
configurations in these two averaged approaches. Figure~\ref{damping} 
demonstrates that the level-to-level study is crucial for the EA process in the 
W$^{27+}$ ion.

\begin{figure}
 \includegraphics[scale=0.4]{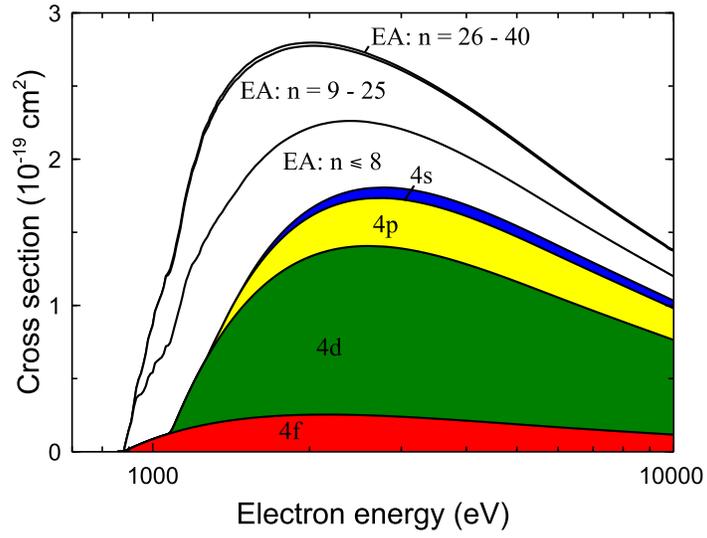}
 \caption{\label{total} (Color online) Total LLDW ionization cross section for 
the W$^{27+}$ ion. The contributions of DI from different shells are highlighted 
by different colors. EA contributions from the excitations to various shells 
are presented. }
\end{figure}

Figure~\ref{total} presents the total ionization cross sections for the direct 
and indirect ionization parts where the indirect part accounts for the radiative 
damping. In this case, the level-to-level calculations are presented. The 
contributions of DI from the $4s$, $4p$, $4d$, and $4f$ shells are highlighted. 
The largest contribution comes from the ionization of the $4d$ electron. As it 
is shown above, the EA channel for the excitations from the $4d$ shell has also 
the largest impact compared to the $4s$ and $4p$ contributions (Fig.~\ref{ea4l}). 
For the DI process, the influence of the $4s$ shell is very small. To ensure our 
data can be easily utilized in modeling studies, we present the total ionization 
cross sections Table \ref{t2}. 

It can be seen that the EA process increases the total cross section at the peak 
nearly by 40 \%  (Fig.~\ref{total}). This contribution would be about two times 
smaller if the excitations to the higher shells ($n \geqslant 9$) were not 
considered. The EA channels which include only the excitations with 
$n \leqslant 8$ and $l \leqslant 3$ provide for the EA process approximately 
10\%  of the total cross section at the peak. 

The ionization cross sections determined for the metastable [Kr]$4d^{10}5s$ 
configuration are approximately by 60 \% higher for the EA process compared to 
the ground configuration when the radiative damping is included for these 
configurations. The contribution of the indirect process from the high-$nl$ 
shells to the total cross sections increases by a factor of two. The DI from 
the $5s$ shell is about two times higher for the metastable configuration 
compared to the ionization from the $4f$ shell of the ground configuration.

\section{Conclusions}

The contribution of the excitations to high-$nl$ shells  has been studied for 
the electron-impact excitation-autoionization process. Calculations for the 
W$^{27+}$ ion illustrate that the excitations to the high-$nl$ shells 
($9 \leqslant n \leqslant 40$) increase the cross section values by a factor 
of two for the indirect part of the ionization process. The largest contribution 
to the indirect part comes from the excitations to $l=4$ orbital. Surprisingly, 
large contribution also arises from the excitations to $l=5$ orbital. The EA 
cross sections for excitations from the $4d$ shell to the shells with 
$n \leqslant 8$ is approximately two times smaller compared to the excitations 
from the $4p$ shell. On the other hand, situation drastically changes for the 
excitations to the higher shells with $9 \leqslant n \leqslant 40$ where the 
excitations from the $4d$ shell dominate. In this case, the EA cross sections 
for the excitations from the $4d$ shell are nearly three times higher. 
The current calculations for the W$^{27+}$ ion show that the radiative damping 
has crucial effect on the cross sections of the indirect process. The cross 
section values decrease by two times for the LLDW results. However, the 
influence of the radiative damping to the subconfiguration-average cross 
sections is approximately 30 \% smaller compared to the influence in the 
level-to-level calculations.

Current results demonstrate that the contribution of the EA channels originating 
from the excitations to the high-$nl$ shells has to be estimated in the analysis 
of the electron-impact ionization process for the highly charged ions. Good 
agreement with experiment for the low-charge states does not ensure that the 
same list of the investigated shells for the EA process is enough when the 
higher-charge states are considered. In order to estimate the contribution from 
the high-$nl$ shells, the convergence of the cross sections of the EA process 
must be checked. 

\section*{Acknowledgement}
This research was funded by European Social Fund under the Global Grant Measure 
(No.: VP1-3.1-\v{S}MM-07-K-02-015). Part of computations was performed on resources 
at the High Performance Computing Center „HPC Saul\.{e}tekis“ in Vilnius University 
Faculty of Physics.

\newpage

\clearpage

\renewcommand{\tabcolsep}{3mm}
\renewcommand{\arraystretch}{1.0}

\begin{table}

\caption{\label{t1} Theoretical lowest energy levels  (in eV) of the 
configurations of W$^{27+}$, W$^{28+}$, and W$^{29+}$ ions. }
 
\begin{tabular}{l c d } 
\hline
\hline 
\multicolumn{1}{c}{ Ion}  & \multicolumn{1}{c}{ Level} & \multicolumn{1}{c}{Energy}   \\  
 \hline 
$W^{27+}$ & $4f(5/2)$  & 0. \\ 
          & $4f(7/2)$     & 3.70    \\
          & $5s(1/2)$     & 139.33  \\
          
$W^{28+}$ & $4d^{10} (0)$   & 878.89  \\
          & $4d_{5/2}^{5}4f_{5/2} (0)$   & 1068.57 \\
          & $4p_{3/2}^{3}4f_{5/2}(1)$   & 1266.25 \\
          & $4s4f_{5/2}(2)$       & 1458.42 \\
          & $3d_{5/2}^{5}4f_{5/2}(0)$   &  2743.71 \\
          
$W^{29+}$ & $4d_{5/2}^{5}({5/2})$ & 2006.64 \\
          & $4d_{3/2}^{3} ({3/2})$ & 2023.13 \\
\hline
\hline

\end{tabular}
\end{table}

\clearpage

\newcolumntype{s}{@{}p{1em}@{}}

\renewcommand{\baselinestretch}{0.9}

\begin{longtable}{rddddsdddd}
 \caption{
\label{t2} 
DI and EA cross sections (in $10^{-20}$ cm$^{2}$) for the  W$^{27+}$ ion. 
DI corresponds to the ionization from the $4s$, $4p$, $4d$, and $4f$ shells. 
The EA contributions from  the $n\le8$, $n\le25$, and $n\le40$ are also 
presented. 
}\\
\hline\hline 
\multicolumn{1}{c}{ Energy }  & 
\multicolumn{4}{c}{ DI} && 
\multicolumn{3}{c}{EA}  &   
\multicolumn{1}{l}{Total}  \\   
\cline{2-5} \cline{7-9}
 \multicolumn{1}{c}{(eV)} & 
\multicolumn{1}{c}{4$s$}  & 
\multicolumn{1}{c}{4$p$}  & 
\multicolumn{1}{c}{4$d$}  & 
\multicolumn{1}{c}{4$f$} && 
\multicolumn{1}{l}{$n$ $\le$ 8}  & 
\multicolumn{1}{l}{$n$ $\le$ 25} & 
\multicolumn{1}{l}{$n$ $\le$ 40} & \\ 
\endfirsthead
\caption[]{ (continued) }  \\
\hline
\hline 
 \multicolumn{1}{c}{ Energy }  & 
\multicolumn{4}{c}{ DI} && 
\multicolumn{3}{c}{EA}  &   
\multicolumn{1}{l}{Total}  \\   
\cline{2-5} \cline{7-9}
 \multicolumn{1}{c}{(eV)} & 
\multicolumn{1}{c}{4$s$}  & 
\multicolumn{1}{c}{4$p$}  & 
\multicolumn{1}{c}{4$d$}  & 
\multicolumn{1}{c}{4$f$} && 
\multicolumn{1}{l}{$n$ $\le$ 8}  & 
\multicolumn{1}{l}{$n$ $\le$ 25} & 
\multicolumn{1}{l}{$n$ $\le$ 40} & \\ 
 \hline 
 \endhead
 \hline 
\multicolumn{10}{r}{{Continued on next page}} \\ 
\endfoot

\hline
880  &      &      &       & 0.010	&& 0.018& 0.018 & 0.018 & 0.028  \\
890  &      &      &       & 0.096	&& 0.701& 0.701 & 0.701 & 0.797  \\
900  &      &      &       & 0.179	&& 1.313& 1.316 & 1.316 & 1.495  \\
930  &      &      &       & 0.409	&& 3.256& 3.920 & 3.920 & 4.329  \\ 
940  &      &      &       & 0.481	&& 3.372& 4.478 & 4.478 & 4.959  \\
950  &      &      &       & 0.549	&& 3.382& 4.817 & 4.817 & 5.366  \\
960  &      &      &       & 0.615	&& 3.515& 5.382 & 5.382 & 5.997  \\
990  &      &      &       & 0.800	&& 4.529& 7.392 & 7.392 & 8.192  \\
1000 &      &      &       & 0.857 && 4.558& 7.798 & 7.798 & 8.655  \\
1010 &      &      &       & 0.913 && 5.140& 8.667 & 8.667 & 9.580  \\
1020 &      &      &       & 0.966 && 5.240& 9.024 & 9.024 & 9.990  \\
1050 &      &      &       & 1.116 && 5.591& 10.146& 10.146& 11.262 \\
1060 &      &      &       & 1.163 && 6.059& 10.821& 10.853& 12.016 \\
1070 &      &      &  0.001& 1.208 && 6.040& 10.862& 11.006& 12.215 \\
1080 &      &      &  0.063& 1.252 && 6.143& 10.988& 11.181& 12.496 \\
1090 &      &      &  0.251& 1.294 && 6.410& 11.224& 11.463& 13.009 \\
1120 &      &      &  1.074& 1.413 && 7.063& 12.143& 12.378& 14.865 \\
1130 &      &      &  1.346& 1.450 && 7.236& 12.439& 12.677& 15.474 \\
1140 &      &      &  1.614& 1.486 && 7.490& 12.964& 13.199& 16.300 \\
1160 &      &      &  2.136& 1.555 && 8.036& 13.756& 13.986& 17.677 \\
1180 &      &      &  2.630& 1.619 && 8.207& 14.256& 14.482& 18.732 \\
1200 &      &      &  3.096& 1.680 && 8.184& 14.668& 14.890& 19.666 \\
1240 &      &      &  3.956& 1.791 && 8.230& 15.400& 15.615& 21.361 \\
1280 &      & 0.051&  4.727& 1.888 && 8.186& 15.619& 15.911& 22.577 \\
1340 &      & 0.374&  5.742& 2.014 && 7.954& 15.677& 15.964& 24.094 \\
1410 &      & 0.817&  6.744& 2.134 && 7.536& 15.262& 15.543& 25.239 \\
1460 & 0.002& 1.107&  7.358& 2.206 && 7.307& 14.874& 15.183& 25.855 \\
1500 & 0.064& 1.311&  7.797& 2.256 && 7.182& 14.558& 14.858& 26.286 \\
1650 & 0.252& 1.918&  9.107& 2.394 && 6.589& 13.305& 13.578& 27.249 \\
1880 & 0.441& 2.518& 10.374& 2.503 && 5.874& 11.820& 12.059& 27.895 \\
2000 & 0.510& 2.731& 10.794& 2.527 && 5.576& 11.185& 11.409& 27.970 \\
2080 & 0.548& 2.845& 11.007& 2.535 && 5.399& 10.810& 11.026& 27.961 \\
2300 & 0.627& 3.077& 11.389& 2.529 && 4.971& 9.927 & 10.124& 27.746 \\
2400 & 0.654& 3.151& 11.487& 2.518 && 4.805& 9.585 & 9.775 & 27.585 \\
2700 & 0.710& 3.294& 11.579& 2.463 && 4.391& 8.742 & 8.914 & 26.959 \\
3000 & 0.742& 3.355& 11.505& 2.389 && 4.068& 8.094 & 8.253 & 26.244 \\
3500 & 0.762& 3.355& 11.158& 2.253 && 3.655& 7.284 & 7.426 & 24.955 \\
3700 & 0.763& 3.334& 10.982& 2.198 && 3.520& 7.020 & 7.157 & 24.435 \\
4100 & 0.758& 3.271& 10.605& 2.091 && 3.280& 6.556 & 6.684 & 23.410 \\
4600 & 0.742& 3.171& 10.122& 1.966 && 3.023& 6.052 & 6.171 & 22.173 \\
6000 & 0.681& 2.861&  8.870& 1.672 && 2.479& 4.981 & 5.080 & 19.163 \\
7000 & 0.637& 2.656&  8.133& 1.511 && 2.199& 4.426 & 4.513 & 17.449 \\
8000 & 0.595& 2.474&  7.506& 1.379 && 1.981& 3.993 & 4.072 & 16.026 \\
9000 & 0.558& 2.311&  6.959& 1.266 && 1.808& 3.653 & 3.725 & 14.819 \\
10000& 0.526& 2.169&  6.481& 1.171 && 1.660& 3.401 & 3.470 & 13.381 \\
                                   
\hline                             
\hline                             
                                   
                                   
\end{longtable}                    

\end{document}